\documentclass[aps,pre,twocolumn,showpacs]{revtex4}

\usepackage{epsfig,latexsym} 
\usepackage{amsmath,amscd,amssymb}

\def\ket#1{|{#1}\rangle}

\begin{document}
 
\title{Quantum algorithm for exact Monte Carlo sampling}
\author{Nicolas Destainville$^{1,2}$, Bertrand Georgeot$^{1,2}$, and 
Olivier Giraud$^{1,2,3}$}
\affiliation{$^1$ Universit\'e de Toulouse; UPS; Laboratoire de
 Physique Th\'eorique (IRSAMC); F-31062 Toulouse, France\\
$^2$ CNRS; LPT (IRSAMC); F-31062 Toulouse, France\\
$^3$ LPTMS, CNRS and Universit\'e Paris-Sud, UMR 8626, B\^atiment 100,
91405 Orsay, France}


\date{9 March 2010}

\begin{abstract}
We build a quantum algorithm which uses the Grover quantum search procedure
in order to sample the exact equilibrium distribution of a wide range of
classical statistical mechanics systems.  The algorithm is based on recently
developed exact Monte Carlo sampling methods, and yields a polynomial gain
compared to classical procedures.
\end{abstract}

\maketitle


The possibility of using quantum mechanics to treat information and perform
computation has attracted great interest
in the recent past (see e.g. \cite{nielsen} for a review).  Quantum algorithms have been devised, which solve computational problems faster
than their classical counterpart, such as the factorization algorithm of Shor \cite{shor}.  However, relatively few problems have
been identified which are amenable to quantum speedup.   While many works have
proposed methods to simulate quantum systems using
a quantum processor (see e.g. \cite{nielsen} and references therein),
fewer have tried to build quantum algorithms to speed up 
classical physical problems \cite{meyer}. In particular,
statistical physics is the source of many computational problems
which have led to great efforts to
develop efficient classical algorithms.  For example, the goal of many
Monte Carlo algorithms is to sample a configuration set $\Omega$ 
from an equilibrium probability distribution $\pi$~\cite{Newman}.
It is therefore important
to explore the possibilities to use quantum computers
in order to speed up such problems.  Some quantum
algorithms have been proposed to approximate the partition functions
of certain statistical physics models \cite{lidar08}, or even 
obtain it exactly in very specific cases \cite{lidarbiham}.
In the very recent past, many works have focused on quantum algorithms 
implementing classical Markov Chain Monte Carlo (MCMC) 
methods through quantum walks \cite{szegedy,santha,barnum,wocjan}.
In general, these methods give a quadratic gain
compared to classical simulations.

Here we consider 
another type of MCMC algorithm recently developed, the ``coupling from the past'' (CFTP) procedure of Propp and Wilson,
which leads
 in finite time to the exact equilibrium distribution \cite{Propp}.  
We propose a quantum algorithm combining this CFTP procedure and 
the quantum search procedure of Grover \cite{grover}, enabling a
quadratic speedup over the classical algorithm, without using quantum walks.
Our method enables to sample the exact
equilibrium distribution in finite time for a wide class of systems,
 while previous algorithms either provide an approximate version 
whose error has to be controlled \cite{lidar08,szegedy,santha,barnum,wocjan}, 
or are restricted to specific models \cite{lidarbiham}. Our algorithm is 
also rather simple compared
to these other methods, while yielding a comparable polynomial gain.

One of the key issues in classical MCMC algorithms is that
they  must be iterated sufficiently many times so that the final state is a ``typical" configuration, in other words has a probability distribution close to the stationary one, $\pi$, independently of the initial state. In order to get close to the correct distribution $\pi$, one should be able to know
when sufficient convergence is achieved.
In a few particular cases, it is possible to calculate analytically
the relaxation time of the algorithm, i.e. the typical time 
needed to reach stationarity. But in practice, estimating or bounding 
relaxation times is a notoriously difficult mathematical 
problem~\cite{aldous,moore,Destain02,Wilson04,Aldous81,Diaconis91} and one has to rely on heuristic
arguments to infer that stationarity has 
approximately been reached. 

An elegant alternative way to circumvent this issue has been proposed by Propp and Wilson in 1996~\cite{Propp,Novotny}. As detailed below, the CFTP technique is a reformulation of the MCMC procedure that generates {\em exact samples}, in the sense that they are exactly distributed according to the stationary distribution $\pi$. Thus successive calls of the algorithm generate totally uncorrelated samples (see below).
The basic idea is to run the Markov chain ``from the past'', from a time $-t$ up to $t=0$. Now suppose that there exists a time $-T$ such that at $t=0$ all states have coalesced (or ``coupled''), i.e. their evolution through the algorithm has led to the same state $x_c$ of the configuration set
$\Omega$. Then any initial
configuration at $t=-\infty$ would lead to the same state $x_c$, which can thus be seen as the result of an infinite time simulation.  The state $x_c$
is consequently 
distributed exactly according to the stationary distribution. The difficulty of the procedure dwells in the necessity to track the evolution of the whole set $\Omega$, whereas in standard Monte Carlo sampling only one state of $\Omega$ is tracked. 

When stored in a computer memory, the configuration set is always finite. Thus we will consider a discrete time Markov chain~\cite{Grimmett} on a finite configuration set $\Omega$ of cardinality $N$. 
Our quantum algorithm consists in replacing the classical evolution of the $N$ states of $\Omega$ by a quantum evolution of a {\em single} quantum state, namely the superposition of the $N$ ones, $\frac1{\sqrt{N}} \sum_x | x \rangle$; then the Grover quantum search procedure is applied on top
of this quantum evolution to find efficiently if the system has
coalesced. 


Let us now detail the classical CFTP algorithm.
Without loss of generality, a state $x \in \Omega$ (with $\Omega$ of cardinal $N$) can be coded by $n$ classical bits $b_i=0,1$ as $x =  b_1\ldots b_n $ and $N\leq2^n$.
 Let $P$ be the transition matrix of the Markov chain. Its elements are the transition probabilities $P(x,y)$, with $P(x,y) \geq 0$ being the conditional
probability that the chain is in the state $x$ at time $t+1$ given that
it was in the state $y$ at time $t$. The chain is supposed to be reversible, which means that it satisfies the
detailed balance condition~\cite{Grimmett,Newman}: there exists a probability distribution on $\Omega$, denoted by $\pi$, such that 
$\pi(x) P(y,x) = \pi(y) P(x,y)$
for all states $x$ and $y$. This condition ensures that $\pi$ is a stationary distribution. We assume that $\pi$ exists and is unique, in which case it coincides with the equilibrium distribution (see \cite{Grimmett} for further details). If $P(x,T | x_0,0)$ is the
probability that the chain is in the state $x$ at time $T$ given that
it was in the state $x_0$ at $t=0$, then~\cite{Grimmett}
\begin{equation}
\lim_{T \rightarrow
\infty} P(x,T | x_0,0) = \pi(x).
\end{equation}

This result is central in traditional Monte Carlo sampling: if the algorithm (the Markov chain) is iterated long enough, then the probability distribution of its final state is close to the stationary one. 

A Monte Carlo step at time $t$ can be
seen as a map $f_t:\Omega \to \Omega$, determined by a
randomly generated parameter $\alpha_t$ as $f_t(\cdot) = \phi(\cdot,\alpha_t)$. Thus once the random numbers $\alpha_t$ are set, each step of
the algorithm is completely deterministic. If $T$ is the duration (number of steps), then the
algorithm is entirely coded by the map $F_T=f_{T} \circ \ldots  \circ f_2
\circ f_{1}$. 
A standard way of performing Monte Carlo sampling consists in following
the dynamics of a single initial state during a
sufficiently large time and averaging a physical observable $\mathcal{O}$
over time iterates.
In this case, the statistical error on the numerical measure of $\langle \mathcal{O} \rangle$ can be estimated using 
the relaxation time $\tau_\mathcal{O}$ of $\mathcal{O}$.  Indeed,
the algorithm behaves 
 as if roughly $T/\tau_\mathcal{O}$ independent
realizations were measured, leading to an error
 ${\rm Err}_\mathcal{O} = \sqrt{2 \tau_\mathcal{O}/T}\Delta \mathcal{O}$, where $T$ is the total simulation duration
and $\Delta \mathcal{O}$ the standard deviation of $\mathcal{O}$~\cite{Newman}.
 In principle, $\tau_\mathcal{O}$ can itself be numerically measured through the correlation function of $\mathcal{O}$, 
$C_\mathcal{O}(s)=\langle \mathcal{O}(t+s)\mathcal{O}(t)\rangle_t-\langle \mathcal{O}\rangle^2 \propto \exp(-s/\tau_\mathcal{O})$. However, $C_\mathcal{O}(s)$ is itself an equilibrium quantity that can be measured only if the system has reached stationarity, and the measured $\tau_\mathcal{O}$ may be only representative of a long transient regime instead of equilibrium. This is particularly critical in disordered, glassy systems where it is impossible to ascertain whether equilibrium has indeed been reached, because the system is likely to get trapped for a long time in the many metastable states~\cite{Bouchaud,Ritort03,Krauth}.

Instead of iterating the states in the future as in the traditional
method explained just above, the 
CFTP procedure goes from the past: we suppose
that we have at our disposal a sequence of random numbers $\alpha_{-1},\alpha_{-2},\ldots$ before starting the algorithm. The CFTP algorithm
constructs the iterates of all the states $x \in \Omega$ until
they have all reached the same state (``coalescence'').  The
essential subroutine of the algorithm [let us call it $\Pi (T)$]
calculates the $N$ computational
paths from time $-T$ up to time $0$ through the map
$G_T=f_{-1}\circ \ldots 
\circ f_{-T}$
and tests at the end whether all histories have coalesced
to the same state.  If the coalescence test fails, the same procedure
is started again from an earlier time.
The algorithm reads:  

\medskip

\noindent $T= 0$ \\
{\tt repeat} \\
\phantom{bla} $T= T+ \Delta T$ (go $\Delta T$ steps back in time) \\ 
\phantom{bla} $\Pi(T)$ (follow all $N$ paths and test coalescence)\\
{\tt until} coalescence is achieved \\


At coalescence, $T$ is such that one has $G_T(x)=x_c$ 
for any $x \in \Omega$.
Thus for any $t' \leq -T$ , $G_{-t'}(x)=x_c$. In particular, $\lim_{t' \rightarrow -\infty} G_{-t'}(x)=x_c$. Therefore $x_c$ can be seen as the result of a Monte Carlo algorithm of infinite duration and is exactly distributed according to $\pi$. It is proven that with probability 1 the algorithm returns a value in finite time~\cite{Propp}.
Successive calls of the algorithm return totally uncorrelated samples,
and $\langle \mathcal{O} \rangle$ is now calculated by averaging 
over realizations instead of time.  The statistical error is now
perfectly controlled: ${\rm Err}_\mathcal{O} = \frac{\Delta \mathcal{O}}{\sqrt{R}}$ where $R$ is the number of realizations 
(independent calls of the algorithm).

The CFTP can be applied to any MCMC problem \cite{proppsoda}. 
In specific instances, it is sufficient to follow the history of a small subset
 of the $N$ states. This is the case
e.g.~if a partial order on $\Omega$ makes it sufficient to follow only 
extremal configurations.
Unfortunately, in general the $N$ states should be followed in parallel, which represents an often prohibitive computational cost. We will propose below a quantum algorithm reducing this cost. 

We denote the average coalescence time of the algorithm by 
$\hat\tau$. To what extent is it related to relaxation times 
$\tau_{\mathcal{O}}$ as discussed above? In principle, $\tau_{\mathcal{O}}$ 
depends on the observable $\mathcal{O}$. But $\tau_{\mathcal{O}}$ is bounded 
above by (and in general on the same order of magnitude as) the 
relaxation time of the Markov chain, denoted by $\tau$ \cite{note1}. This latter
time measures 
the speed of convergence of the probability distribution to $\pi$ and
is equal to the inverse of the first gap of the transition matrix $P$~\cite{Aldous81,Diaconis91}. 
Furthermore, $\tau$ is itself bounded above by (and in general on the 
same order of magnitude as) $\hat\tau$~\cite{Aldous81}, which 
makes CFTP-type techniques so useful to estimate convergence rates, even 
on theoretical grounds. All in all, generally speaking, $\hat\tau \sim 
\tau \sim \tau_{\mathcal{O}}$.  Running the algorithm yields
an accurate estimate of $\hat\tau$ and thus of relaxation times.



\begin{figure}[h]
\begin{center}
\includegraphics[width=.94\linewidth]{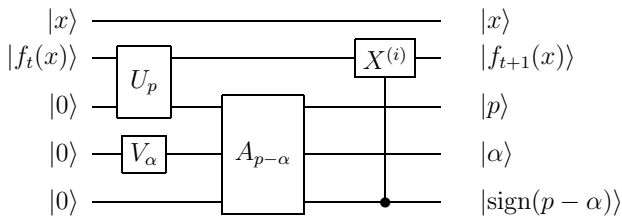}
\end{center}
\vglue -0.3cm
\caption{Circuit for one step of the Monte-Carlo algorithm. 
The unitary operator
  $U_p$ calculates probability $p$, $V_{\alpha}$
is the operator implementing the  random numbers $\alpha$,
and $A_{p-\alpha}$ is a modified adder circuit that gives the sign of
$p-\alpha$.}
\vglue -0.2cm
\label{bitflip}
\end{figure}

Let us now turn to our quantum algorithm.
The essential subroutine
$\Pi(T)$ of the classical CFTP algorithm follows the history of the
$N$ states $x\in\Omega$ and tests coalescence of all states. 
The
quantum algorithm will start from a register holding the $N$ initial states
$\ket{x}$, $0\leq x\leq N-1$, coded on $n$ qubits, and follow each history in
parallel. A second register holds the results of the
successive application of the maps $f_t$. Calculation of the iterates is done
with the help of ancilla registers. To illustrate the computational
scheme, we will first specialize our presentation to the case of the Ising
model with Glauber dynamics. In this case
the observable $\mathcal{O}$ could be e.g. the magnetization. Starting from the totally separable state,  Hadamard
gates are applied to each qubit to put the register into an equal
superposition of all states. Suppose that
after $t$ iterations the quantum state reads
\begin{equation}
\sum_{x=0}^{N-1}\ket{x}\ket{H_{t,T}(x)}\ket{0}\ket{0}\ket{0},
\end{equation}
where $H_{t,T}= f_{-T+t-1}\circ \ldots  \circ f_{-T}$.
The next Monte Carlo step consists in flipping a certain spin $i$
with some probability function of the energy difference $\Delta E$ between 
configuration $H_{t,T}(x)$ and the same configuration with spin $i$ flipped. 
This spin $i$ may be chosen at random; alternatively one
can consider each spin one after another (sequential sweep). The spin 
is flipped 
with probability 
$p=[1+\exp(\beta \Delta E)]^{-1}$
or left unchanged with probability $1-p$. Here $\beta$ is the inverse
temperature. 
In terms of quantum registers, one has to evaluate the energies
associated with configurations $\ket{y}=\ket{H_{t,T}(x)}$ and 
 $\ket{y^{(i)}}=X^{(i)}\ket{y}$
(where the Pauli matrix $X^{(i)}$ flips spin $i$), by 
arithmetic operations controlled by 
the second register. The probability $p$
 is then calculated on the third register
and the random number $\alpha_t$, uniformly
distributed on [0,1], is put on the fourth register. 
The sign of $\alpha_t - p$ is computed, and the one-qubit fifth 
register $\ket{s}$ is set to
$\ket{0}$ if $\alpha-p \geq 0$ and $\ket{1}$ if $\alpha-p<0$.
The bit-flip matrix $X^{(i)}$ is applied, controlled by $\ket{s}$. 
The last three registers are then reset to $|0\rangle$ in the usual way
by running the operations backwards.
The circuit in Fig.~\ref{bitflip} shows one step of the iteration
algorithm before reset of these registers.

After $T$ steps the quantum state reads
\begin{equation}
\label{superp}
\sum_{x=0}^{N-1}\ket{x}\ket{G_T(x)}.
\end{equation}
If $T$ is such that all $G_T(x)$ are equal then the histories have coalesced
and a measure of the second register yields  an element in $\Omega$
distributed exactly according to the stationary distribution $\pi$. Since all iterations are
performed in parallel this step requires an average time $\hat{\tau}$. The crucial point in
the CFTP algorithm is that the exact distribution is obtained if and only  if
all $G_T(x)$ are equal. Consider for instance the case where 
the  $G_T(x)$ take two different values, say $y_1$
and $y_2$. Then one might detect that the states have not coalesced onto a
unique value as soon as one obtains
different results after measuring the second register
upon repeated runs of the procedure (with the same random numbers). If the
states have not coalesced then the process has to be restarted from
an earlier time. Obviously, only in the case where the probabilities of 
measuring $y_1$ and $y_2$ are both high will different 
outcomes be obtained quickly upon measurement. In the
extreme case where there is a unique $x_0\in\Omega$ such that $G_T(x_0)=y_2$
while all other $x$ verify $G_T(x)=y_1$, almost all measurements of the
second register will give $y_1$, and the state will be almost
indistinguishable from the state where all $G_T(x)$ are equal. Since the CFTP
algorithm requires to distinguish these cases, the idea is to use the Grover algorithm to
amplify the probability amplitude of the unknown 
noncoalesced state $\ket{x_0}$, so
that it can be detected. If we first measure the second register and
consistently get the value $y_1$ then our aim is to detect whether all $x$
verify $G_T(x)=y_1$ or not. Since we know the value of $y_1$, 
we can attach to our quantum state a one-qubit register in the state
$\ket{z}=\frac{1}{\sqrt{2}}(\ket{0}-\ket{1})$. We then perform 
the operation 
$\ket{x}\ket{G_T(x)}\ket{z}\mapsto\ket{x}\ket{G_T(x)}\ket{z+\varphi(x)}$, where
$\varphi(x)=0$ if $G_T(x)=y_1$ and $\varphi(x)=1$ otherwise, and erase 
all the registers but the first one. This gives a phase e$^{i\pi}$ 
to states which
do not verify $G_T(x)=y_1$. One can thus apply Grover
iterations to magnify the amplitude of
the noncoalesced states.
The whole sequence above corresponds to one
``oracle'' step of the Grover iteration, and has to be performed
using the same random numbers $\alpha_t$.  

What is the speed-up on $\Pi(T)$ obtained by this
procedure?
 Suppose that after
$T$ time steps $M$  states $x$ are not coalesced. We consider the case
$M\ll N$, since otherwise noncoalescence is easily detected by a few
measurements or even classically. Then $O(\sqrt{N/M})$ Grover iterations 
are required 
(even though $M$ is unknown \cite{counting}), each 
using $O(T)$ operations. The total number of quantum gates in this
case is $\sim T\sqrt{N/M}$.  As the random numbers 
can be generated classically once and for all beforehand, 
their computational cost
is $\sim T$ and thus negligible.

As explained above, the complete algorithm proceeds by performing a
certain number of calls of $\Pi(T)$ until coalescence. 
Let us evaluate the speedup of the quantum algorithm for $\Pi(T)$
compared to the classical procedure.  We consider that preliminary runs
enable quickly to estimate a suitable $\Delta T \lesssim \hat{\tau}$ 
which is gradually improved.
To iterate classically the Monte Carlo
steps on one computational path will cost a certain number of 
computational operations $g(N)$. Let us first suppose that the dynamics 
is rapid, with a short coalescence time $\hat{\tau}$ with  
$g(N)\sim \ln^a N$.  
This means that the dynamics is polynomial in the physical
system size $n$.  The classical algorithm will need to follow $\sim N/M$ 
computational paths to detect the ones which did not coalesce. The
total cost is $\sim g(N) N/M \sim (N \ln^a N) /M $.  In contrast,
in the case of the quantum algorithm, $O(\sqrt{N/M})$ calls to the oracle
are required, and the oracle performs the $N$ evolutions in parallel
in also $\sim g(N)\sim \ln^a N$ operations. The total speedup of
the quantum algorithm will be $O(\sqrt{N/M})$, a quadratic gain.
If the dynamics is torpid, with a long coalescence time
$\hat{\tau}$ with  $g(N)\sim N^c$, then the classical algorithm requires
$\sim N^{c+1}/M $ operations whereas the quantum one costs only 
$\sim N^{c+1/2}/\sqrt{M} $ elementary steps.
The relative gain gets smaller for increasing $c$,
going from quadratic for small $c$ to almost zero for
$c\rightarrow \infty$. 

The estimates above assume that all
initial states, stored in a $n$-bit register and coded
by all numbers $x$ between $0$ and $N-1$, are physically admissible. This is
the case e.g. for spin problems.
However,  in many interesting cases one must
inspect a large number of states ($N=2^n$), of which only a
small subset ($N_a=N^{b}$, $b <1$) are admissible.  
This is the case e.g. for dimer, spanning tree or
hard core lattice gas problems.  In this situation, 
one can use a modified version of the above quantum algorithm.
Indeed, starting from an equal
superposition of the $N$ states, one can build in the Grover oracle [before
each $\Pi(T)$ step] a subroutine
which recognizes the nonadmissible states, and overwrites in this case 
the second register  with a known admissible state $x_a$, so that the
latter is used as an initial state in the dynamics.  In this case,  
the quantum algorithm for $M= O(1)$ still requires $\sim g(N) \sqrt{N}$
steps, since the Grover search is applied on the whole Hilbert space
of dimension $N$.  To obtain both the equilibrium distribution
and the relaxation time, the classical algorithm needs 
$\sim N$ operations to identify admissible states, and
$\sim g(N) N^{b}$ operations to run the CFTP on the admissible states.
If $g(N)\sim \ln^a N$, the gain is unchanged. If $g(N)\sim N^c$, it
is $\max(c+b,1)/(c+1/2)$.  The gain
is polynomial in all cases except for $b <1/2$ and $c>1/2$
(very long relaxation time and very few admissible states).

The algorithm proposed here presents several advantages compared to the
recently proposed method for simulating Markov Chain systems
\cite{szegedy,santha,barnum,wocjan}.  
These procedures use quantum walks to approximate
the stationary distribution, in a time typically
quadratically faster than the classical convergence time. The speedup over
classical computation is therefore comparable, but, in our case, 
we obtain a sampling of the exact 
stationary distribution
rather than an approximate version of it with errors that 
have to be controlled.  Our quantum algorithm is also
simpler.  Another advantage of our method is that the relaxation time
is directly related to the average coalescence time \cite{Aldous81} and thus 
can be accurately measured.  

It has been proven that the Grover algorithm is
optimal, in the sense that the number of calls to the oracle cannot
be lower (see \cite{nielsen} and references therein).  Our quantum algorithm 
 can therefore be
improved only by speeding up the oracle part.
This may be possible by combining our algorithm with 
techniques used in the other approaches of
\cite{lidarbiham,lidar08,szegedy,santha,barnum,wocjan}.
As the algorithm developed here is very general,
applying to a wide class of systems without any structure taken into account,
tailored algorithms may achieve a larger gain in specific cases.



\begin{thebibliography}{99}
\bibitem{note1} $\tau_\mathcal{O}=\tau$ if the average value of $\mathcal{O}$
on the first eigenmode of $P$ is nonvanishing; otherwise 
$\tau_\mathcal{O} <\tau$.
\end{thebibliography}
\end{document}